\documentclass[fleqn,usenatbib]{mnras}

\usepackage{newtxtext,newtxmath}

\usepackage[T1]{fontenc}

\DeclareRobustCommand{\VAN}[3]{#2}
\let\VANthebibliography\thebibliography
\def\thebibliography{\DeclareRobustCommand{\VAN}[3]{##3}\VANthebibliography}

\usepackage{graphicx}	
\usepackage{amsmath}	
\usepackage{subfigure, color}
\usepackage{xcolor}
\usepackage{cleveref}
\hypersetup{colorlinks,allcolors=black}

\title[AU~Mic and Prox~Cen Galactic cosmic ray fluxes]{The strong suppression of Galactic cosmic rays reaching AU~Mic\,b, c and Prox~Cen\,b} 

\author[A. L. Mesquita et al.]{
A. L. Mesquita$^{1}$\thanks{E-mail:mesquita@strw.leidenuniv.nl},
D. Rodgers-Lee$^{2,3}$, 
A. A. Vidotto$^{1}$ and 
R. D. Kavanagh$^{1}$
\\
$^{1}$Leiden Observatory, Leiden University, PO Box 9513, 2300 RA, Leiden, The Netherlands\\
$^{2}$School of Physics, Trinity College Dublin, The University of Dublin, D02 E8C0, Ireland\\
$^{3}$School of Cosmic Physics, Dublin Institute for Advanced Studies, 31 Fitzwilliam Place, D02 XF86, Ireland
}

\date{Accepted XXX. Received YYY; in original form ZZZ}

\pubyear{2022}

\begin{document}
\label{firstpage}
\pagerange{\pageref{firstpage}--\pageref{lastpage}}
\maketitle

\begin{abstract}
The propagation of Galactic cosmic rays is well understood in the context of the solar system but is poorly studied for M~dwarf systems. Quantifying the flux of cosmic rays reaching exoplanets is important since cosmic rays are relevant in the context of life. Here, we calculate the Galactic cosmic ray fluxes in AU~Mic and Prox~Cen planetary systems. We propagate the Galactic cosmic rays using a 1D cosmic ray transport model. We find for Prox~Cen\,b, AU~Mic\,b and AU~Mic\,c that the Galactic cosmic ray fluxes are strongly suppressed and are lower than the fluxes reaching Earth. We include in our models, for the first time for a star other than the Sun, the effect of radial particle drift due to gradients and curvatures in the stellar magnetic field. For Prox~Cen we find that the inclusion of particle drift leads to less suppression of Galactic cosmic rays fluxes than when it is excluded from the model. In the case of AU~Mic we explore two different wind environments, with a low and high stellar wind mass-loss rate. For AU~Mic, the particle drift also leads to less suppression of the Galactic cosmic ray fluxes but it is only significant for the high mass-loss rate scenario. However, both wind scenarios for AU~Mic suppress the Galactic cosmic rays strongly. Overall, careful modelling of stellar winds is needed to calculate the Galactic cosmic ray fluxes reaching exoplanets. The results found here can be used to interpret future exoplanet atmosphere observations and in atmospheric models.
\end{abstract}




\section{Introduction}
In a series of works, we have investigated the propagation of Galactic cosmic rays in M dwarfs \citep{Mesquita2021, Mesquita2022}, with the aim to derive the flux of cosmic rays that reach the habitable zone of these stars. We also calculated the radiation dose reaching an exoplanet surface \citep{Mesquita2022}. Here, we continue this series of works, focusing on two nearby M dwarfs whose orbiting planets are amenable to atmospheric characterisation: AU~Mic and Prox~Cen. M dwarfs have habitable zones that are close-in due to their low luminosity. Because of our current observational biases towards finding close-in planets, this means that it is easiest to observe planets orbiting in the habitable zone of M dwarfs. However, being in the habitable zone \citep[i.e. where surface liquid water can exist,][]{Kasting1993, Selsis2007} is not thought to be the only element necessary for a planet to sustain life. Many factors can influence exoplanet habitability \citep[see e.g.][]{Meadows2018}. One important factor for habitability is cosmic rays. 

Cosmic rays are a source of ionisation that can drive the production of prebiotic molecules in Earth-like \citep{Airapetian2016} and Jupiter-like atmospheres \citep{Barth2020}. Hence, cosmic rays may have been important for the beginning of life on Earth and other planets. Fingerprint ions, such as H$_3$O$^+$, H$_3^{+}$ and NH$_4^+$ \citep{Helling2019,Barth2020}, have been identified as a good indication of ionisation in hot gas giant atmospheres due to both Galactic and stellar cosmic rays (also known as stellar energetic particles). Spectroscopic observations with the James Webb Space Telescope \citep[JWST,][]{Gardner2006} and the Atmospheric Remote-sensing Infrared Exoplanet Large-survey \citep[ARIEL,][]{Tinetti2021} of absorption features due to these fingerprint ions may be able to constrain the cosmic ray fluxes in exoplanet atmospheres in the future. In order to interpret future exoplanet atmosphere observations and also to provide input for atmospheric chemistry models, it is necessary to know the flux of Galactic cosmic rays reaching exoplanets.

As Galactic cosmic rays propagate through stellar systems they are suppressed, in a energy-dependent way, by the existence of a magnetised stellar wind \citep[e.g.][]{Griebmeier2005, Sadovski2018, Herbst2020, Mesquita2021, Rodgers2021-2, Mesquita2022}. This is analogous to what occurs in the solar system, known as the modulation of Galactic cosmic rays \citep[e.g.][]{Potgieter2013}. The Galactic cosmic ray fluxes reaching an exoplanet are affected by diffusion, advection, adiabatic losses and particle drift processes that depend on the stellar wind properties (the magnetic field and velocity). They are also affected by the Galactic cosmic ray spectrum in the interstellar medium (ISM) and the ISM properties (velocity and density). 

Particle drifts, caused by gradients and curvatures of the heliospheric magnetic field, have been pointed out as an important ingredient in the description of cosmic ray transport inside the heliosphere \citep{Jokipii1977}. Many works have included the effect of particle drifts to describe the transport of cosmic rays in the heliosphere \citep{Strauss2012, Strauss2014, Potgieter2013, Vos2015, Potgieter2017, Kopp2021}, in good agreement with observations at Earth. Particle drifts were also included to study the young Sun system \citep{Cohen2012}. However, for other stars the effect of particle drift are usually neglected \citep{Griebmeier2005,Herbst2020, Mesquita2021, Rodgers2021-2, Mesquita2022}. One reason for this is that to implement particle drift it is necessary to know the 3D large-scale stellar magnetic field geometry. At this point, the 3D large scale stellar magnetic field geometry is known for more than a hundred stars \citep[e.g.][]{Vidotto2014, See2015, klein2021, Klein2021-b}, albeit the stellar wind mass-loss rates have been determined for a smaller number of stars \citep[e.g.][]{Wood2004, Wood2021}. Since different stars have different winds due to their different (and some times extreme) magnetic fields and high rotation rates, 3D models of the stellar wind are an important tool to study the propagation of cosmic rays and the effects of particle drift in stellar systems.

In this work, we investigate the well-studied M dwarfs AU~Mic and Prox~Cen as they both host detected exoplanets \citep{Anglada2016, Plavchan2020, Martioli2021}. Prox~Cen\,b is an Earth-size planet orbiting at 0.048\,au \citep{Anglada2016}. AU~Mic\,b and AU~Mic\,c are both Neptune-size planets at 0.064\,au and 0.11\,au, respectively \citep{Plavchan2020, Martioli2021}. In particular, Prox~Cen\,b orbits in the habitable zone and AU~Mic is a JWST target which will be observed soon using the Near InfraRed Camera (NIRCam) to search for undetected close-in planets. In addition, the large scale magnetic field geometry has been reconstructed with Zeeman-Doppler Imaging (ZDI) for both Prox~Cen and AU~Mic \citep{klein2021, Klein2021-b}. In terms of their stellar wind properties, Prox~Cen has a mass-loss rate constrained by Lyman-$\alpha$ observations to be $< 4 \times 10^{-15}\,M_{\sun}$~yr$^{-1}$ \citep{Wood2001}. While mass-loss rate constraints for AU~Mic are less stringent, models predict that it ranges from 10 to 1000 times the solar wind mass-loss rate \citep{Plavchan2020, Chiang2017}. \citet{Kavanagh2021} modelled the winds of the Prox~Cen and AU~Mic planetary systems using these observational constraints for the stellar magnetic fields and mass-loss rates.

In this paper, we use the 3D stellar wind model results from \citet{Kavanagh2021} to compute the propagation of Galactic cosmic rays through the AU~Mic and Prox~Cen astrospheres (analogous to the heliosphere). We use our 1D model of cosmic ray transport \citep{Rodgers2020, Mesquita2021, Rodgers2021-2, Mesquita2022} to calculate the spectrum of Galactic cosmic rays within the astrospheres. Here we implement in this model the effect of particle drift for the first time in an M dwarf system. This paper is structured as follows: our modelling framework is explained in \Cref{sec:models}. We then apply our model to Prox~Cen (\Cref{sec:prox-results}) and AU~Mic (\Cref{sec:aumic-results}), where we compute the cosmic ray fluxes that reach the known planets of these systems, as well as in their habitable zones. In the case of AU~Mic, we also explore different assumptions for the stellar wind properties, in order to investigate how a poorly constrained mass-loss rate affects our results. \Cref{sec:conclusions} presents a discussion of our results and conclusions.

\section{Modelling Framework}
\label{sec:models}
Our modelling framework consists of the following ingredients. First, we adopt a stellar wind model to derive the stellar wind properties (described in \Cref{sec:winds}). Second, using the stellar wind properties, we model the cosmic ray transport as a diffusive-advective process as they propagate through the stellar wind (see \Cref{sec:crs}). The Galactic cosmic rays are injected at the boundary between the stellar wind and the ISM. We determine this outer boundary by calculating the size of the astrosphere (see \Cref{sec:ast}).

\subsection{Stellar wind properties}
\label{sec:winds}
In order to study the propagation of Galactic cosmic rays through an astrosphere it is essential to know the stellar wind velocity and magnetic field strength. For AU~Mic and Prox~Cen, we use the 3D magnetohydrodynamic (MHD) wind simulations from \citet{Kavanagh2021}. These simulations are conducted using the AWSoM model \citep{Holst2014, Jin2017}, which was originally derived for the study of the solar wind. Additionally, to benchmark our cosmic ray transport code with observations of cosmic rays at the Earth, we also use a model of the solar wind, which is shown in \Cref{sec:sun}.

In their study, \citet{Kavanagh2021} used reconstructed surface magnetic field maps for Prox~Cen and AU~Mic that were obtained using the ZDI method \citep{klein2021, Klein2021-b}. In the AWSoM model, the stellar wind is heated and dissipated by Alfvén waves injected at the base of the chromosphere. To account for the range of mass-loss rates expected for AU~Mic, \citet{Kavanagh2021} computed two different stellar wind models: one with $\dot{M}=5.5 \times 10^{-13}\,M_{\sun}$~yr$^{-1}$ and the other with $\dot{M}=1.2 \times 10^{-11}\,M_{\sun}$~yr$^{-1}$. We refer to these models as `AU~Mic low' and `AU~Mic high', respectively. Although these two stellar wind models have the same stellar wind magnetic field map, they have different Alfvén wave fluxes, giving rise to different stellar wind mass-loss rates and terminal velocities. 

\begin{figure*}
	\includegraphics[width=.8\paperwidth]{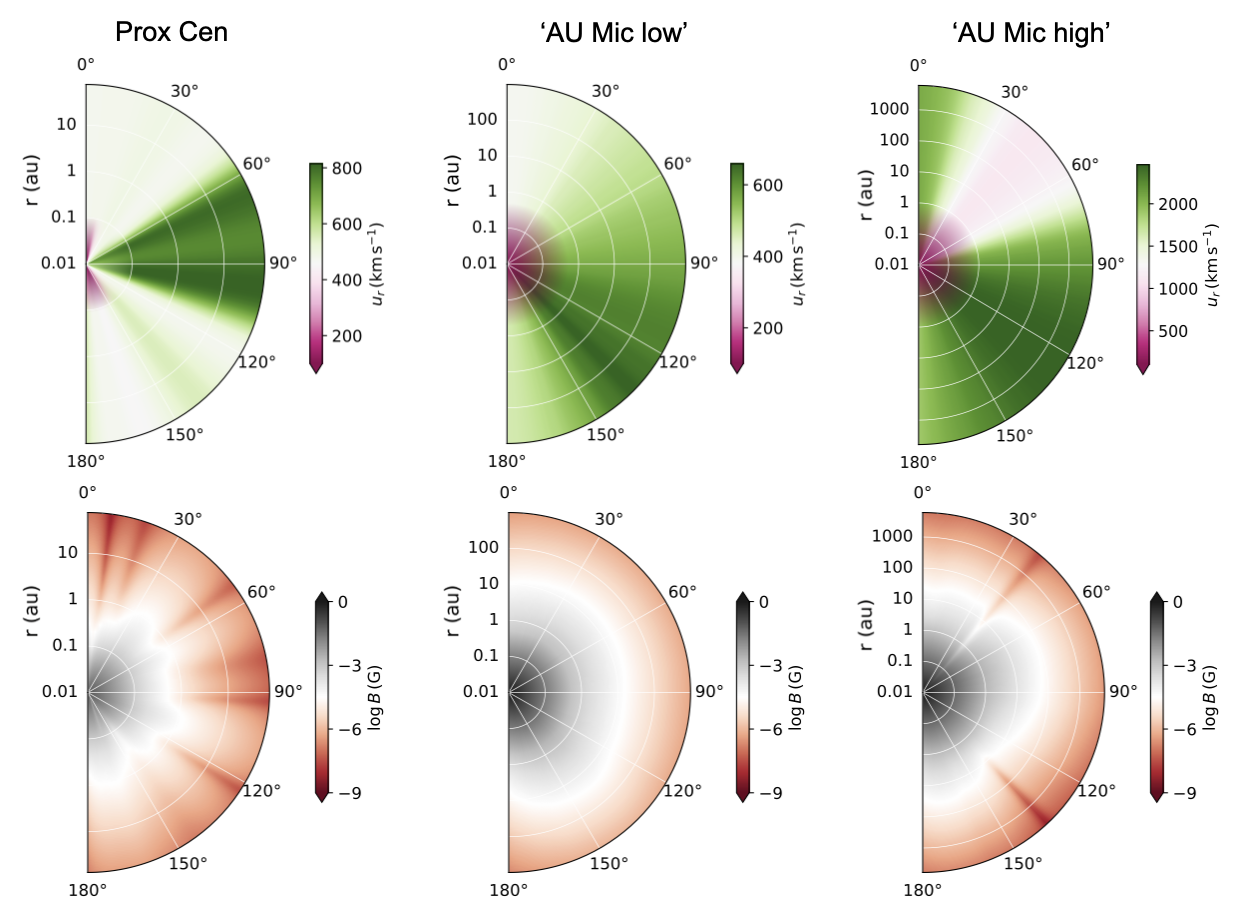}
    \caption{Equatorial plane cut of the stellar wind properties, from left to right, for Prox~Cen, `AU~Mic low' and `AU~Mic high'. The top row shows the stellar wind velocity and the bottom row shows the magnitude of the total magnetic field for the full radial extent of each star's astrosphere. The values shown for small orbital distances are from  3D stellar wind models \citep{Kavanagh2021} and the values shown at large distances are an extrapolation (see \Cref{sec:extrapolation} for details).}
    \label{fig:wind}
\end{figure*}

\Cref{tab:stars} summarises some of the properties for the stars studied here, including their adopted stellar wind properties. \Cref{fig:wind} shows the equatorial plane cut of the stellar wind velocity (top row) and magnetic field intensity (bottom row) for Prox~Cen, `AU~Mic low' and `AU~Mic high'. The values shown for small orbital distances ($r<100\,R_{\star}$) are from the 3D stellar wind model of \citet{Kavanagh2021}. Beyond that, the values shown are  extrapolated values (see \Cref{sec:extrapolation} for details). The stellar magnetic field geometry shapes the stellar wind velocity distribution, i.e., a smoother magnetic field geometry results in a smoother velocity. For instance, Prox~Cen shows a complex magnetic field geometry (more than simply a dipole component) and streams with very different speeds. AU~Mic instead shows a less complex magnetic field geometry (relatively close to a dipole) and consequently less streams with different speeds. Compared to our model for the Sun, with $u_\infty =540$\,km~s$^{-1}$, (see \Cref{sec:sun}), Prox~Cen (810\,km~s$^{-1}$) and `AU~Mic low' (650\,km~s$^{-1}$) have comparable terminal wind velocities while `AU~Mic high' has a much faster terminal velocity of 2460\,km~s$^{-1}$. In relation to the stellar surface magnetic field, both Prox~Cen and AU~Mic have a much higher magnetic field intensity in comparison to our model for the Sun, $B \sim 2$\,G (see \Cref{sec:sun}).

\begin{table*}
 \caption{Properties of the Prox~Cen and AU~Mic stellar systems. The quantities without reference are calculated in this work.}
 \begin{tabular}{ccccc}
  \hline
   Physical parameter & Prox~Cen & \multicolumn{2}{c}{AU~Mic} & Unit\\
    &  & `AU~Mic low' & `AU~Mic high' &  \\
  \hline
  Stellar mass ($M_{\star}$) & 0.12$^a$ & \multicolumn{2}{c}{0.50$^b$} & $M_{\sun}$\\
  Stellar radius ($R_{\star}$) & 0.14$^a$ & \multicolumn{2}{c}{0.75$^b$} & $R_{\sun}$\\
  Distance ($d$) & 1.30$^a$ & \multicolumn{2}{c}{9.79$^b$} & pc\\
  Habitable zone & 0.03 -- 0.09$^c$ & \multicolumn{2}{c}{0.26 -- 0.66$^c$} & au\\
  Average stellar surface magnetic field ($B_{\star}$) & 200$^d$ & \multicolumn{2}{c}{475$^e$} & G\\
  ISM/star relative velocity ($v_{\rm ISM}$) & 25$^f$ & \multicolumn{2}{c}{19} & km~s$^{-1}$\\
  ISM ram pressure ($P_{\rm ISM}$) & $2.5 \times 10^{-12}$ & \multicolumn{2}{c}{$1.5 \times 10^{-12}$} & dyn~cm$^{-2}$ \\
  Stellar wind mass-loss rate$^g$ ($\dot{M}$) & $5.0\times 10^{-15}$ & $5.5\times 10^{-13}$ & $1.2\times 10^{-11}$ & $M_{\sun}$~yr$^{-1}$\\
  Stellar wind ram pressure ($P_{\rm ram}$) & $3.3 \times 10^{-6}$ & $1.2\times 10^{-5}$ & $4.6 \times 10^{-4}$ & dyn~cm$^{-2}$\\
  Terminal velocity$^g$ ($u_\infty$) & 810 & 650 & 2460 & km~s$^{-1}$\\
  Astrospheric size ($R_{\rm ast}$) & 75 & 980 & 6140 & au\\
  \hline
 \end{tabular}
 \label{tab:stars}
 \newline
  $^a$\citet{Anglada2016}; $^b$\citet{Plavchan2020}; $^c$calculated using the prescription of \citet{Kopparapu2014}; $^d$\citet{klein2021}; $^e$\citet{Klein2021-b}; $^f$\citet{Wood2001}; $^g$\citet{Kavanagh2021}.
\end{table*}

\subsection{The astrosphere size}
\label{sec:ast}
The astrosphere is the `bubble' region around a star dominated by its stellar wind. The astrosphere size can be very important for the propagation of Galactic cosmic rays because outside this region cosmic rays are not suppressed. The size of the astrosphere can be calculated as
\begin{equation}
    R_\text{ast}=\sqrt{\frac{P_{\text{ram}}(R)}{P_{\text{ISM}}}}R,
    \label{eq:astro}
\end{equation}
where $P_{\text{ram}}$ is the ram pressure of the stellar wind expressed as
\begin{equation}
P_{\text{ram}}=\rho u^2=\frac{\dot{M} u}{4\pi R^2},
\end{equation}
where $u$ is the wind velocity, $R$ is a given reference distance beyond which the wind has reached its terminal velocity. The ISM ram pressure, $P_{\text{ISM}}$, can be expressed as
\begin{equation}
P_{\text{ISM}}=m_p n_{\rm ISM}\nu_{\rm ISM}^2,
\end{equation}
where $m_p$ is the proton mass, $n_{\text{ISM}}$ is the total ISM number density of hydrogen and $\nu_{\text{ISM}}$ is the ISM velocity as observed by the star.

To derive the total ISM number density of hydrogen ($n_{\rm ISM}$), we use the same ISM neutral hydrogen density ($n_{\rm n}=0.14\,\text{cm}^{-3}$) and ionised hydrogen density ($n_{\rm i}=0.1\,\text{cm}^{-3}$) as given by Model 10 of \citet{Wood2000} which is a good fit for nearby systems in the local ISM. Assuming that the ISM cloud along the line of sight towards AU~Mic is most consistent with the Mic cloud, we derive the heliocentric flow vector in the direction of AU~Mic and the radial velocity of the Mic cloud using the ISM Kinematic Calculator \citep{Redfield2008}. The heliocentric velocity of AU~Mic is calculated using its proper motion \citep{Gaia2020} and its radial velocity \citep{Fouque2018}. From these assumptions, we calculate the ISM velocity and pressure as seen by AU~Mic to be $v_{\rm ISM}=19$\,km~s$^{-1}$ and $P_{\rm ISM}=1.5 \times 10^{-12}$dyn~cm$^{-2}$. If instead we assume that the ISM cloud along the line of sight towards AU~Mic is consistent with the Local Interstellar Cloud (LIC), the ISM velocity does not vary significantly ($v_{\rm ISM}=19.4$\,km~s$^{-1}$ ). For Prox~Cen, we use the ISM velocity as given by \citet{Wood2001} and shown in \Cref{tab:stars}. The astrospheric size of Prox~Cen and AU~Mic (for the two stellar wind scenarios) are given in \Cref{tab:stars}. The astrospheric size is proportional to the stellar wind mass-loss rate. Compared with the heliosphere, where $R_{\rm ast}=122$\,au  \citep[as observed by {\it Voyager 1,}][]{Stone2013,Stone2019}, Prox~Cen's astrosphere is 40\% smaller while $R_{\rm ast}$ for `AU~Mic low' is 8 times larger and for `AU~Mic high' it is 50 times larger. 

\subsection{Cosmic ray transport equation}
\label{sec:crs}
When interacting with magnetised stellar winds, Galactic cosmic rays can experience global changes in their intensity and energy, referred to as modulation. The cosmic ray propagation through a magnetised stellar wind can be described by the transport equation of \citet{Parker1965}. We use a 1D cosmic ray transport model to calculate the cosmic rays fluxes within the M dwarf systems. The model used here is based on the model described in \citet{Rodgers2020} and was already used for M dwarfs \citep{Mesquita2021, Mesquita2022}. Here, we add for the first time, for a star other than the Sun, the effect of particle drift.

We numerically solve the time-dependent, spherically symmetric transport equation for cosmic rays, as given by
\begin{equation}
    \frac{\partial f}{\partial t}=\nabla\cdot(\mathbf{\kappa} \nabla f) - (\mathbf{u} +  \langle \mathsf{\mathbf{v}}_d \rangle )\cdot \nabla f+\frac{1}{3}(\nabla\cdot \mathbf{u})\frac{\partial f}{\partial \ln p},
    \label{eq:transport}
\end{equation}
where $f$ and $p$ are the cosmic ray phase space density and momentum, respectively. The diffusion of cosmic rays is represented by the first term on the right-hand side of \Cref{eq:transport} and depends on the diffusion coefficient, $\kappa$. The stellar wind velocity, $\mathbf{u}$, and the particle drift velocity, $\langle \mathsf{\mathbf{v}}_d \rangle$, appear in the second term and represent advection of the cosmic rays. The last term represents momentum advection, also known as adiabatic losses, which moves the cosmic rays to lower energies. More details of the model can be found in \citet{Rodgers2020}. Our results in \Cref{sec:prox-results,sec:aumic-results} present the differential intensity of cosmic rays, $j$, rather than the phase space density, $f$, where $j(T)=p^2f(p)$. $T$ is the cosmic ray kinetic energy.

We use logarithmically spaced spatial and momentum grids. The inner spatial boundary is set as 0.01\,au and the outer boundary is set as the astrosphere size, $R_{\rm ast}$, for each star (see Table\,\ref{tab:stars}). The spatial grid includes the orbits of the known exoplanets and the habitable zones. The spatial grid has 80 grid zones for Prox~Cen, 100 grid zones for the Sun and `AU~Mic low' and 120 grid zones for `AU~Mic high'. The spatial resolution varies for each system because they have different astrosphere sizes. The inner spatial boundary condition is reflective and the outer spatial boundary condition is fixed as the Local Interstellar Spectrum (LIS) for Galactic cosmic rays. The LIS that we adopt is given by the model fit to the {\it Voyager 1} observations using Eq. (1) of \citet{Vos2015} \footnote{As discussed in \citet{Mesquita2021, Mesquita2022}, this is a good assumption since according to $\gamma$-ray observations the inferred cosmic ray spectrum within a region of 1\,kpc in the local Galaxy is consistent with {\it Voyager} measurements in the local ISM \citep{Neronov2017}.}. The LIS is considered to be constant as a function of time in our simulations. The momentum grid ranges from $p_{\rm min}=0.15$\,GeV$/c$ to $p_{\rm max}=100$\,GeV$/c$ with 60 momentum bins. The limits of the momentum grid are chosen due to the fact that cosmic rays with higher energies than the upper limit are quite uncommon and not important for planetary atmosphere chemistry \citep{Rimmer2013}, while cosmic rays with lower energy than the lower limit are not expected to reach the planetary surface \citep{Atri2017}. Low-energy particles, however, can deposit energy in the planetary atmosphere \citep{Rodgers2020} and may be relevant for chemical models \citep[e.g.][]{Griebmeier2015, Barth2020}. Both, inner and outer, momentum boundary conditions are outflow boundary conditions. 

\subsubsection{The diffusion coefficient}
\label{sec:diffusion}
The diffusion coefficient, $\kappa$, in \Cref{eq:transport}, depends on the level of turbulence in the stellar magnetic field and the cosmic ray momentum. From quasi-linear theory \citep{Jokipii1966, Schlickeiser1989}, the diffusion coefficient can be denoted as
\begin{equation}
  \frac{\kappa(r,p)}{\beta c}=\eta_0\left(\frac{p}{p_0}\right)^{1-\gamma}r_\text{L},
\end{equation}
where $r_\text{L}$ is the cosmic ray Larmor radius, $p_0=3\,$GeV$/c$ and $\beta=v/c$ is the ratio between the cosmic ray's velocity and the speed of light. $\eta_0$ represents the level of turbulence present in the system. In our simulations we set $\eta_0=1$ to be consistent with other works \citep{Rodgers2020, Mesquita2021, Mesquita2022}. $\gamma$ determines the power-law dependency with momentum. The type of turbulence can be described by different prescriptions, such as, Bohm-type ($\gamma=1$), Kolmogorov-type ($\gamma=5/3$) and MHD-driven ($\gamma=3/2$) turbulence. We note that the Galactic cosmic ray spectrum calculated at any given distance is affected by the type of turbulence adopted. The type of turbulence present in M dwarf systems is still unknown and here we simply adopt $\gamma=1$, i.e. Bohm-type turbulence. This has been used frequently in many works \citep{Svensmark2006,Cohen2012, Rodgers2020, Mesquita2021, Rodgers2021-2, Mesquita2022} and fits the present-day observations at Earth quite well \citep{Rodgers2020}.

The magnetic field and velocity are important wind properties in the context of Galactic cosmic ray propagation. For instance, the stronger the magnetic field, the more the Galactic cosmic ray spectrum is suppressed (for fixed values of $\gamma$ and $\eta_0$). This is because a strong magnetic field implies smaller diffusion coefficients. This means that generally advective processes become more important, resulting in more modulation of cosmic rays. The same is valid for the wind velocity, however to a lesser extent \citep{Mesquita2022}, i.e. a stronger stellar wind velocity results in more advection which suppresses the cosmic ray fluxes. The stellar wind density does not attenuate the cosmic rays, as the density is very low. However, it influences the size of the astrosphere (see \Cref{eq:astro}). In cases where advection is the dominant physical process, the size of the astrosphere can affect the Galactic cosmic ray fluxes \citep{Rodgers2021-2}.

\subsubsection{The particle drift}
The averaged particle drift velocity, $ \langle \mathsf{\mathbf{v}}_d \rangle$, is caused by gradients and curvatures in the stellar magnetic field. The drift velocity can be expressed as a function of the magnetic field as \citep[][]{Jokipii1993}
\begin{equation}
    \langle \mathsf{\mathbf{v}}_d \rangle = \frac{p\beta c^2}{3e} \mathbf{\nabla} \times \frac{\mathbf{B} }{B^2},
\end{equation}
where $B$ is the stellar wind magnetic field. The drift velocity also depends on the cosmic ray momentum $p$ and velocity $v$ (through $\beta$). 

In our 1D propagation model, we only include the radial component of the drift velocity, which can be expressed in spherical coordinates as
\begin{equation}
    \langle \mathsf{\mathbf{v}}_d \rangle = \frac{p\beta c^2}{3e}\frac{1}{r \sin\theta} \left(\underbrace{\frac{\partial}{\partial \theta}\left(\sin\theta\frac{B_\phi}{B^2}\right)}_{\rm i} - \underbrace{ \frac{\partial}{\partial \phi}\left(\frac{B_\theta}{B^2}\right)}_{\rm ii} \right ) \mathbf{e}_r,
    \label{eq:drift-terms}
\end{equation}
where $\theta$ is the polar angle ($\theta=90^\circ$ at the equator) and $\phi$ is the azimuthal angle.

\begin{figure}
	\includegraphics[width=\columnwidth]{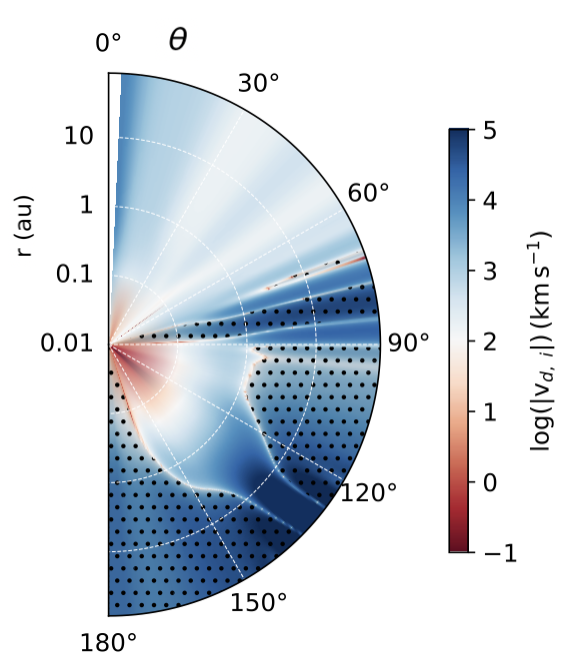}
	\includegraphics[width=\columnwidth]{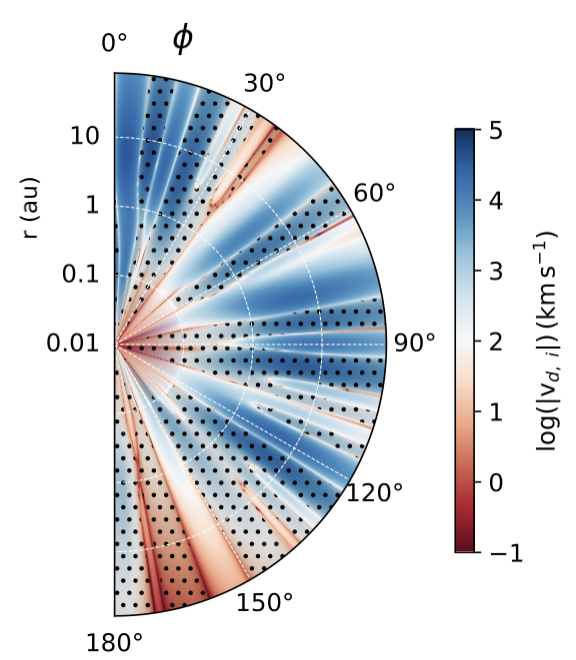}
    \caption{Components of the radial drift velocity for Prox~Cen for cosmic rays with 1\,GeV/c momentum. Top panel: term i of \Cref{eq:drift-terms} as a function of $r$ and $\theta$ for $\phi$ direction where the stellar wind is fastest (see \Cref{fig:wind} top row). Bottom panel: term ii of \Cref{eq:drift-terms} as a function of $r$ and $\phi$ for the equatorial plane. In both plots the dotted regions represent regions with negative velocities.}
    \label{fig:drift}
\end{figure}

\Cref{fig:drift} shows the absolute value of the drift velocity components that we calculate for Prox~Cen for cosmic rays with $p=\,$1\,GeV/c. The top panel shows term i of \Cref{eq:drift-terms} as a function of radial distance and polar angle, $\theta$, for the value of $\phi$ where the stellar wind velocity have its strongest intensity ($\phi=97^\circ$). The bottom panel shows term ii of \Cref{eq:drift-terms} as a function of radial distance and the azimuthal angle, $\phi$, for the equatorial plane ($\theta=90^\circ$). The dotted areas represent regions with negative velocities. The momentum dependency of the drift velocity is a multiplicative factor, i.e., the drift velocity is 100 times higher for 100\,GeV/c momentum cosmic rays than for 1\,GeV/c cosmic rays. Both terms of the drift are spatially non-linear and vary as a function of $r$, $\theta$ and $\phi$ significantly, as shown in \Cref{fig:drift}. In the top panel of \Cref{fig:drift} we did not include $\theta=0^\circ$ because $1/\sin\theta$ in \Cref{eq:drift-terms} implies that the drift value becomes infinite at this point. In our simulations we only include the drift velocity as a function of $r$ and $p$ for a single value of $\theta$ and $\phi$ where i and ii are combined. We will discuss the effect of the 2D/3D spatial variation of the drift velocity in \Cref{sec:conclusions}. 

The effects of $\langle \mathsf{\mathbf{v}}_d \rangle$ have been neglected in past studies of M dwarf systems. However, if $\langle \mathsf{\mathbf{v}}_d \rangle$ is larger or comparable to the stellar wind velocity, $u$, it can affect the modulation of cosmic rays. The particle drift acts as an additional advective term. Hence, when the particle drift velocity is positive, i.e. radially outward pointing, then the Galactic cosmic ray fluxes are reduced. We will discuss this in more detail in \Cref{sec:prox-results,sec:aumic-results}.

\section{Prox~Cen: the Galactic cosmic ray fluxes including particle drift effects}
\label{sec:prox-results}
Prox~Cen is the closest star to the solar system and has a terrestrial planet in its habitable zone. Here, we investigate the propagation of Galactic cosmic rays within its astrosphere. Since we use a 1D cosmic ray transport model, we select particular cuts of the 3D wind simulations to calculate the drift velocity at any given orbital distance. We investigate the Galactic cosmic ray fluxes at the equatorial plane ($\theta=90^\circ$) because this allows us to investigate the effect of Galactic cosmic rays at the planet's equatorial orbit. Since the stellar wind velocity also varies with $\phi$, we further fix the value of $\phi=97^\circ$ for Prox~Cen. We note that at the chosen $\theta$ and $\phi$ the stellar wind has already reached its terminal velocity within the 3D stellar wind simulation domain.

\begin{figure}
	\includegraphics[width=\columnwidth]{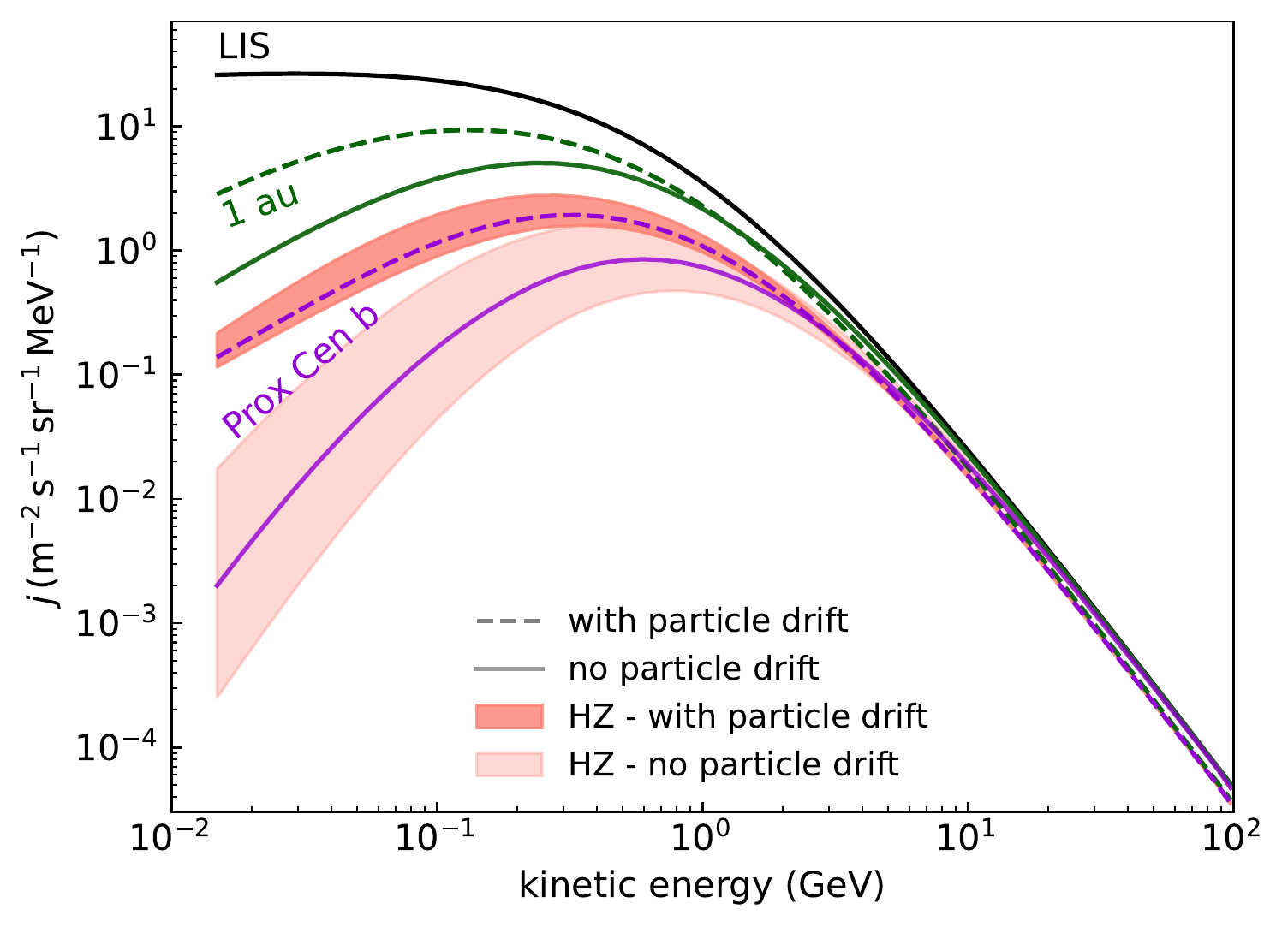}
    \caption{Differential intensity of Galactic cosmic rays as a function of cosmic ray kinetic energy for Prox~Cen. The solid lines represents the Galactic cosmic ray flux without considering the particle drift and the dashed lines when the drift is considered. The black solid line is the LIS. The green lines are the Galactic cosmic ray flux at 1\,au and the violet lines are the Galactic cosmic ray flux at Prox~Cen b. The shaded areas are the Galactic cosmic ray flux in the habitable zone, both accounting for (dark salmon) and neglecting (light salmon) particle drift. In the case of Prox~Cen, the inclusion of particle drift leads to Galactic cosmic ray fluxes that are an order of magnitude higher for cosmic ray energies $\lesssim 0.1$ GeV.}
    \label{fig:prox}
\end{figure}

\Cref{fig:prox} shows the intensity of cosmic rays as a function of kinetic energy for Prox~Cen when particle drift is considered (dashed lines) and when it is neglected (solid lines) at different distances. The shaded areas represents the Galactic cosmic ray flux in the habitable zone when particle drift is considered (dark salmon) and when it is neglected (light salmon). The black line is the LIS. When particle drift is included, Prox~Cen shows a considerably higher flux of cosmic rays for particles with kinetic energy below 1\,GeV. At the orbit of Prox~Cen\,b, the intensity of 0.1\,GeV energy cosmic rays is 7 times higher when drift is considered (dashed violet line) in comparison with results when drift is neglected (solid violet line). A similar trend is observed at 1\,au (green lines) and in the habitable zone (shaded areas). 

\begin{figure}
	\includegraphics[width=\columnwidth]{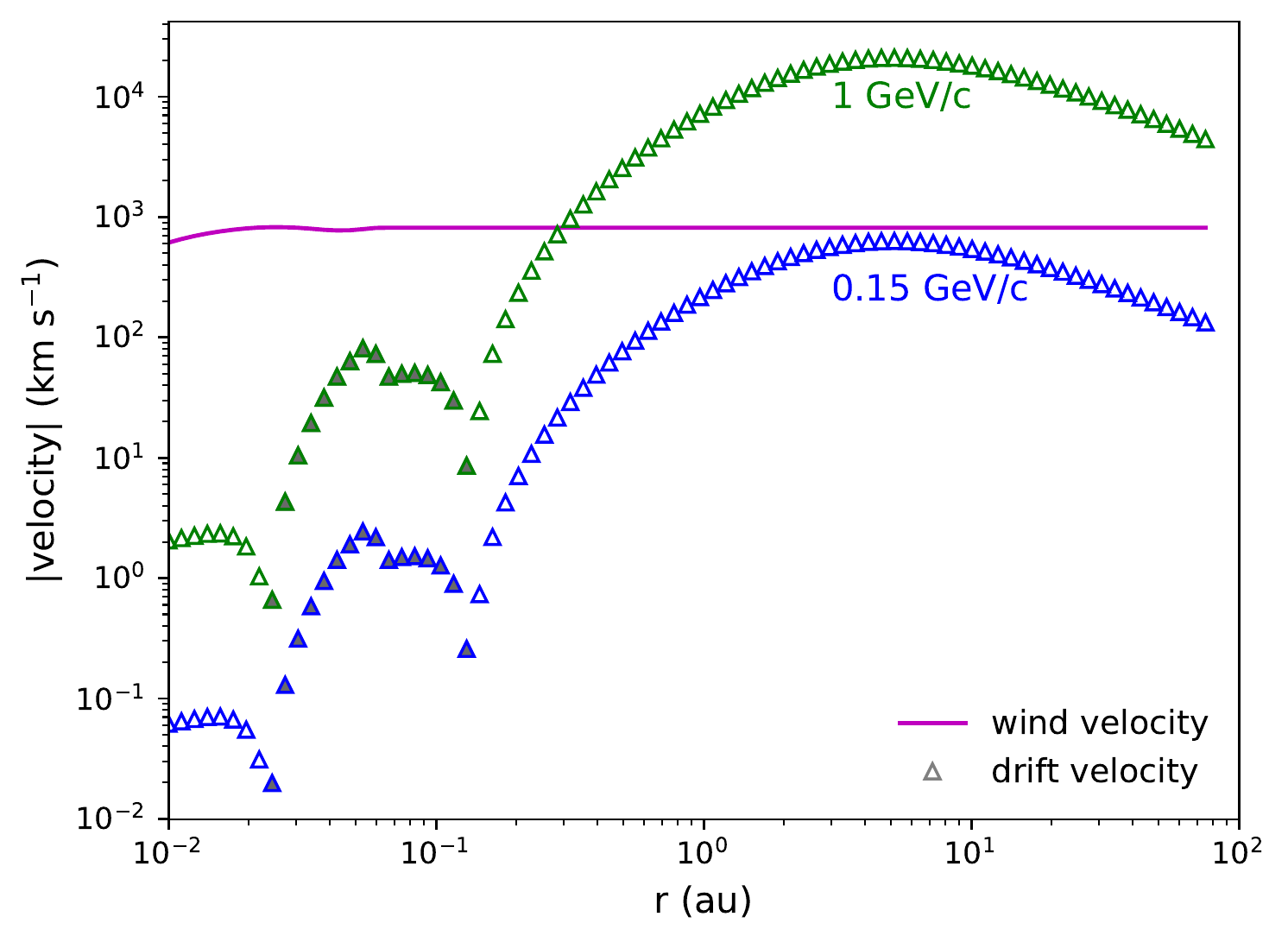}
    \caption{Stellar wind and radial drift velocity as a function of distance for Prox~Cen at $\theta=90^{\circ}$, $\phi=97^{\circ}$. The magenta line is the stellar wind velocity from \citet{Kavanagh2021}. The green and blue triangles are the absolute value of the radial drift velocity for cosmic rays with 1\,GeV/c and 0.15\,GeV/c momentum, respectively. Open triangles represent negative values and filled triangles represent positive values. Cosmic rays with higher momenta will have larger drift velocity values and vice-versa.}
    \label{fig:drift-prox}
\end{figure}

The effects we see with the inclusion of particle drifts can be understood by comparing the drift velocity with the stellar wind velocity. \Cref{fig:drift-prox} shows the stellar wind velocity and the drift velocity profiles (for $\theta=90^{\circ}$, $\phi=97^{\circ}$) for Prox~Cen. The green triangles are $|\langle \mathsf{v}_d \rangle|$ for cosmic rays with $p=1$\,GeV/c, while the blue triangles are for $p=0.15$\,GeV/c. The stellar wind velocity is lower than $|\langle \mathsf{v}_d \rangle|$ for $r>0.3\,$au (green triangles) and, as a consequence, the drift velocity affects the modulation of cosmic rays. The Galactic cosmic ray flux within the Prox~Cen system is higher when particle drifts are included because the drift velocity is large and negative in value (open triangles in \Cref{fig:drift-prox}). If, on the other hand, the drift velocity were positive, the cosmic ray flux would be reduced. 

Our results likely depend on the direction we choose to study ($\theta$ and $\phi$ values). In the case of Prox~Cen for instance, it is possible that for azimuthal angles where the wind velocity is lower the particle drift may play an even more important role in modulating the cosmic rays. In \Cref{fig:drift} we can see that each term of the drift velocity varies significantly with $\theta$ and $\phi$. A 2D or 3D cosmic ray simulation would be necessary to investigate the cosmic ray propagation in other directions.

\section{AU~Mic: the effect of an unconstrained mass-loss rate on Galactic cosmic ray fluxes}
\label{sec:aumic-results}
AU~Mic is a young M dwarf stellar system that hosts two Neptune-size exoplanets. Both exoplanets have close-in orbits and are not within the habitable zone. As discussed in \Cref{sec:winds}, we investigate Galactic cosmic ray propagation for the two stellar wind scenarios explored by \citet{Kavanagh2021}.  Although their magnetic field profiles are similar, the mass-loss rate and terminal velocity of `AU~Mic high' are, respectively, 21 and 4 times larger than the values for `AU~Mic low'. Because of their different wind properties, the calculated sizes of the astrosphere differ for these two wind models. Similar to Prox~Cen, we investigate the flux of Galactic cosmic rays for AU~Mic in the equatorial plane and we set $\phi=133^\circ$ for `AU~Mic low' and $\phi=126^\circ$ for `AU~Mic high'. 

\begin{figure}
	\includegraphics[width=\columnwidth]{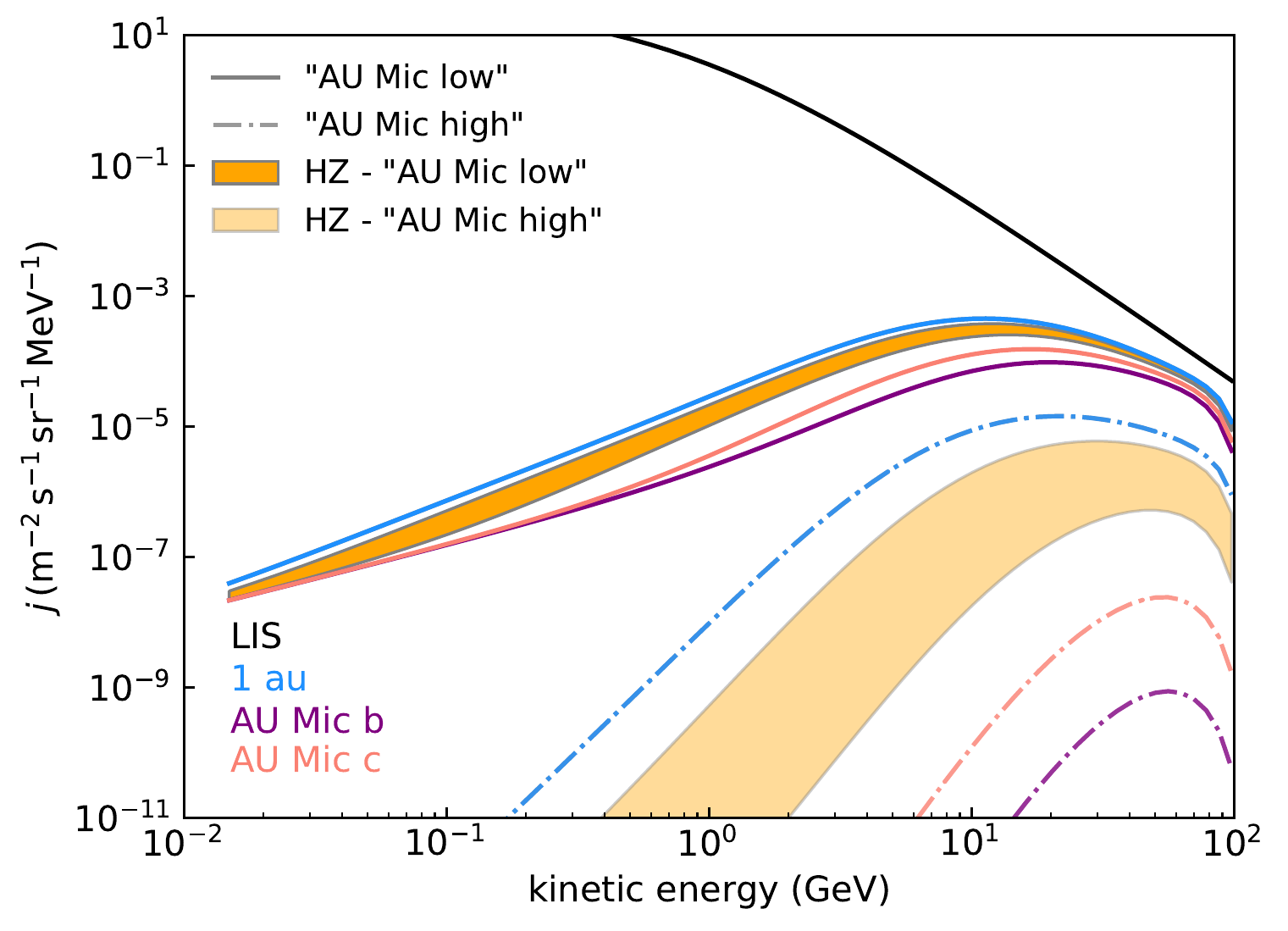}
    \caption{Differential intensity of Galactic cosmic rays as a function of cosmic ray kinetic energy for `AU~Mic low' (solid lines) and `AU~Mic high' (dash dotted lines). Blue, purple and salmon lines are the differential intensity of cosmic rays at 1\,au, AU~Mic\,b and AU~Mic\,c, respectively. The orange shaded areas represent the fluxes in the habitable zone, where the dark orange is for `AU~Mic low' and the light orange for `AU~Mic high'. The stellar wind properties for the `AU~Mic high' case are more efficient at suppressing the Galactic cosmic ray fluxes.}
    \label{fig:aumic}
\end{figure}

\Cref{fig:aumic} shows the cosmic ray fluxes as a function of kinetic energy for the two wind scenarios of AU~Mic. Here, in this figure, we only show the results of our calculations that consider particle drift. The orange shaded areas represents the cosmic ray fluxes in the habitable zone, where the dark orange is for `AU~Mic low' and the light orange for `AU~Mic high'. Blue, purple and salmon lines are the cosmic ray fluxes at 1\,au, AU~Mic\,b and AU~Mic\,c. In general, the Galactic cosmic ray fluxes for `AU~Mic high' are much reduced in comparison to `AU~Mic low' (see for instance blue lines in \Cref{fig:aumic}). The difference is particularly sharp for cosmic rays with energy lower than 1\,GeV. Within each wind prescription, the intensity of cosmic rays at AU~Mic\,b and AU~Mic\,c is similar since the planets have close orbits. 

In the case of AU~Mic, the unconstrained mass-loss rate of the stellar wind and the different terminal velocities from each model strongly affect the Galactic cosmic ray intensities within the system as shown in \Cref{fig:aumic}. This is opposite to what was derived for GJ~436 by \citet{Mesquita2021}, in which the wind properties did not strongly affect the cosmic ray fluxes. The reason for this, is that AU~Mic is dominated by advective processes (with or without the inclusion of particle drift), while GJ~436 is dominated by diffusive processes. When advection dominates it results in a stronger suppression of cosmic rays, while when diffusion dominates it results in little (or no) modulation. The diffusive/advective processes are correlated with the stellar wind velocity and magnetic field (see \Cref{sec:diffusion} and the diffusive/advective time-scales in Section 3.2 of \citealt{Mesquita2021}). In addition, when advection dominates the size of the astrosphere can affect the suppression of cosmic rays \citep{Rodgers2021-2}.

In the case of `AU~Mic low', the drift velocity profile for cosmic rays with 1\,GeV/c momentum is negative but almost negligible when compared with the stellar wind velocity profile and, as a consequence, the cosmic ray fluxes are not influenced by the inclusion of drift. For `AU~Mic high', the drift velocity for cosmic rays with 1\,GeV/c momentum is negligible ($\lesssim 30\,\rm km~s^{-1}$) for $r<1\,$au. For $1\,\rm au<r<10\,\rm au$, it is positive and relatively slow, while for $r>10\,$au $|\langle \mathsf{v}_d \rangle|$ is negative and fast (but slower than the drift velocity intensities observed for Prox~Cen). Thus, the Galactic cosmic ray intensities are slightly affected by the inclusion of particle drift for close-in distances ($r<1\,$au). For large distances ($r>1\,$au) neglecting the drift can underestimate the cosmic ray fluxes. For instance, we find at 744\,au for 1\,GeV energy cosmic rays,  the intensity of Galactic cosmic rays when drift is considered is two orders of magnitude higher when compared with no drift. 

Overall, particle drift is an important ingredient in cosmic ray modulation, particularly when $|\langle \mathsf{v}_d \rangle|>u$. Here, we fixed $\phi$ and $\theta$ values as previously discussed. However, we can expect that the presence of particle drift will affect the cosmic ray modulation of Prox~Cen and `AU~Mic high' at any given location of the equatorial plane. This is due to the fact that particle drift changes the modulation of cosmic ray even in the direction where the wind have the strongest velocity.

\section{Discussion \& Conclusions}
\label{sec:conclusions}
In this paper, we investigated the intensity of Galactic cosmic rays within the astrospheres of two planetary systems: AU~Mic and Prox~Cen. AU~Mic hosts two known close-in Neptune-size exoplanets. Prox~Cen is our closest star and has a known Earth-size exoplanet orbiting within the habitable zone. We used a 1D cosmic ray transport model which includes the effect of particle drift to calculate the cosmic ray fluxes reaching the exoplanets and the habitable zones. The radial particle drift velocity is calculated using 3D MHD stellar wind models for Prox~Cen, AU~Mic (with a `low' and `high' mass loss rate) and the Sun for comparison \citep[][see also \Cref{sec:sun}]{Kavanagh2021}. This is the first time that radial drift velocities have been quantified for M dwarf systems.

Overall, the inclusion of particle drift velocities, due to gradients and curvatures of the stellar magnetic field, has a significant effect on the propagation of Galactic cosmic rays. In general, a negative drift velocity results in a larger intensity of cosmic rays and a positive drift velocity results in a reduction of cosmic rays. This is because the particle drift acts simply as an extra advective process in the transport model. In the two systems studied here, the drift velocity is negative at the cut where we investigated the flux of Galactic cosmic rays. Prox~Cen and `AU~Mic high' both have a larger negative drift velocity while for `AU~Mic low' the drift velocity is rather slow for cosmic rays with 1\,GeV/c momentum. Consequently, the modulation of cosmic rays is significantly affected by the inclusion of drift for Prox~Cen and `AU~Mic high' but only slightly affected in the case of `AU~Mic low'.

For Prox Cen, at 1\,au (green dashed line in \Cref{fig:prox}) the Galactic cosmic ray fluxes are larger in comparison to Earth. This is because Prox~Cen's magnetic field profile is weaker at larger distances (e.g., from 1\,au to the astrosphere edge) in comparison to the solar system and results in less suppression of Galactic cosmic rays for the same distance. In contrast, both AU~Mic scenarios (low and high) have a much stronger stellar magnetic field profile in comparison to the solar system, resulting in an intense suppression of Galactic cosmic rays at the same distance.

\citet{Herbst2020} have also investigated the Galactic cosmic ray fluxes at Prox~Cen\,b, albeit without considering the effects of particle drift. They found that cosmic rays were not effectively reduced in comparison to the LIS. In contrast, our results show lower cosmic ray fluxes at Prox~Cen\,b. For instance, compared to \citet{Herbst2020}, the cosmic ray fluxes at Prox~Cen\,b with energy 1\,GeV is 7 times smaller in our simulations when particle drift is not included and 5 times smaller when particle drift is included. The differences in our results and \citet{Herbst2020} is likely to be related to the different wind and ISM properties used and the cosmic ray transport properties. They used a surface magnetic field strength of $\sim 600\,$G measured from Zeeman broadening \citep{Reiners2008} and a wind velocity of $1500\,\rm km~s^{-1}$ based on MHD simulations \citep{Garraffo2016}. Both quantities are higher than the values we used. We also calculate different astrosphere sizes, 122\,au \citep{Herbst2020} versus 75\,au in our work, as a result of our different wind and ISM properties. If the drift velocities for Prox~Cen used in our work were included in \citet{Herbst2020} work, they would probably find an even larger flux of Galactic cosmic rays than the values they calculate for Prox~Cen\,b.

In the case of AU~Mic, we showed that our lack of a strong constraint for the wind properties (mass-loss rate and wind velocity) strongly affects the intensity of Galactic cosmic rays calculated for the system and the size of the astrosphere (980--6140\,au). This is in contrast with the results found by \citet{Mesquita2021} for GJ~436, in which the lack of knowledge of $u$ and $\dot{M}$ did not strongly affect the flux of Galactic cosmic rays in the system. This is related to the physical process that is more dominant \citep{Rodgers2021-2}. While GJ~436 is dominated by diffusion which leads to little (or no) modulation of cosmic rays, AU~Mic is dominated by advection processes which strongly suppress cosmic rays. Interestingly, AU~Mic is dominated by advective processes even without the inclusion of the particle drift. Additionally, because `AU~Mic high' has a larger astrosphere size the Galactic cosmic rays have further to travel through a region of the stellar wind dominated by advection in comparison with `AU~Mic low'.

In addition to drift effects, the stellar magnetic field and the wind velocity are important ingredients to effectively modulate the Galactic cosmic ray fluxes throughout the astrosphere. A strong stellar magnetic field, leading to smaller diffusion coefficients, combined with a fast stellar wind velocity is effective at suppressing the propagation of Galactic cosmic rays inside the astrosphere. The magnetic field intensity can change if the star has an activity cycle, similar to the solar cycle \citep[see e.g.][]{Hathaway2010}. When the Sun is at activity minimum the intensity of Galactic cosmic rays is higher than at activity maximum \citep{Potgieter2013}, with  more than an order of magnitude difference in flux, depending on the cosmic ray energy \citep{Vos2015}. Prox~Cen, similarly to the Sun, has been suggested to have a 7 year activity cycle \citep[e.g.][]{Yadav2016}. The magnetic map adopted in our wind model was derived near its activity maximum \citep{klein2021}. This could imply that during its activity minimum Prox~Cen would likely experience a higher flux of Galactic cosmic rays than the results presented here. In addition, the particle drift will also vary with activity cycle since it depends on the magnetic field geometry.

Another important aspect to consider is that stars themselves can accelerate stellar cosmic rays during events such as flares and coronal mass ejections \citep{Rodgers2021}. Prox~Cen and AU~Mic are both active stars with strong surface magnetic fields \citep{klein2021,Klein2021-b} and frequent flaring activity \citep{Gilbert2021,Gilbert2022} and, as such, they should be efficient at accelerating stellar cosmic rays. As a result, for these stars, stellar cosmic ray fluxes will dominate over Galactic cosmic rays up to a certain energy. Some works have explored the effect of such events for the exoplanets in the AU~Mic and Prox~Cen systems. \citet{Scheucher2020} showed that a strong stellar cosmic ray event would be able to heat the otherwise cold planet Prox~Cen\,b, playing an important role in the planet's habitability. Recently, \citet{carolan2020} investigated the effects of strong stellar winds in the atmospheric evaporation of AU~Mic\,b. They concluded that, even when atmospheric erosion by the stellar wind is not significant, the geometry and ionisation of the escaping atmosphere can change substantially, impacting the interpretation/prediction of spectroscopic transit observations. In addition to strong winds, coronal mass ejections and flares can also affect planetary atmospheres, by increasing atmospheric erosion \citep{Khodachenko2007, hazra2020, hazra2022}. Although these processes are episodic, they are expected to be more frequent and more energetic in young and/or active stars \citep{Aarnio2014}, such as AU Mic. 
 
To better account for the full 3D nature of planetary systems, 2D/3D cosmic ray transport models should be used, similar to models implemented for the solar system. Although the radial drift velocity varies with the polar and azimuthal angles, which we have neglected, it is possible with 2D/3D simulations that diffusion could smooth these variations, resulting in a more continuous distribution of cosmic rays. In this case, the differential intensity might not be too different than the ones we calculated with our model. The polar and azimuthal drift velocities may also act in a similar way.

Finally, the Galactic cosmic rays fluxes at the exoplanets' orbits can be used to further investigate the intensity of Galactic cosmic rays in planetary atmospheres and in atmospheric chemistry models. Our models can also be used to understand future observations of exoplanet atmospheres with instruments such as JWST and ARIEL.
  
\section*{Acknowledgements}
This project has received funding from the European Research Council (ERC) under the European Union's Horizon 2020 research and innovation programme (grant agreement No 817540, ASTROFLOW). DRL would like to acknowledge that this publication has emanated from research conducted with the financial support of Science Foundation Ireland under Grant number 21/PATH-S/9339. The authors wish to acknowledge the SFI/HEA Irish Centre for High-End Computing (ICHEC) for the provision of computational facilities and support.

\section*{Data Availability}
The data described in this article will be shared on reasonable request to the corresponding author.



\bibliographystyle{mnras}
\bibliography{bib/reference.bib}



\appendix
\section{Galactic cosmic ray propagation in the solar system}
\label{sec:sun}
In order to benchmark our code we calculate Galactic cosmic ray propagation through the heliosphere. Here we explain the solar wind model and the results we obtain at Earth.

To model the solar wind plasma, we follow the same methodologies as in \citet{Kavanagh2021}. Unlike AU~Mic and Prox~Cen, the mass-loss rate of the solar wind has been measured in-situ to be $2\times10^{-14}$\,$M_{\sun}$~yr$^{-1}$ \citep{Cohen2011}. The mass-loss rate of the wind is sensitive to the value used for the Alfvén wave flux per unit magnetic field strength $S_\textrm{A}/B$ in the chromosphere \citep{Boro2020, Kavanagh2021}. For the Sun, this value has been estimated directly from observations to be $1.1\times10^5$\,erg~s$^{-1}$~cm$^{-2}$~G$^{-1}$ \citep{Sokolov2013}. 

\begin{figure}
	\includegraphics[width=\columnwidth]{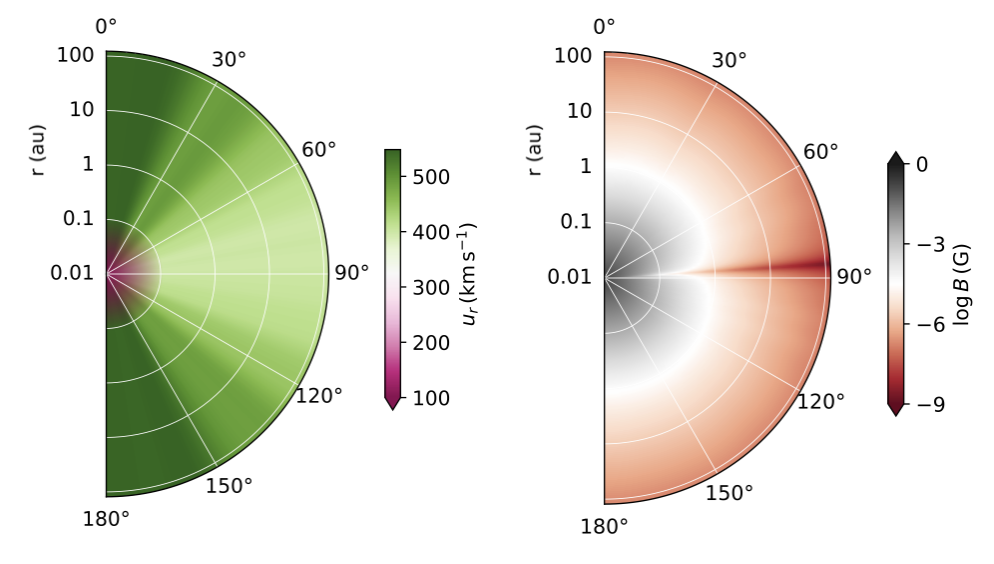}
    \caption{{\it Left}: Equatorial plane cut of the solar wind velocity. {\it Right}: Equatorial plane cut of the magnitude of the total magnetic field for the full radial extent of the heliosphere. The values shown for small orbital distances are from the 3D solar wind model and the values shown at large distances are an extrapolation (see \Cref{sec:extrapolation} for details).}
    \label{fig:wind-sun}
\end{figure}

\begin{figure}
	\includegraphics[width=\columnwidth]{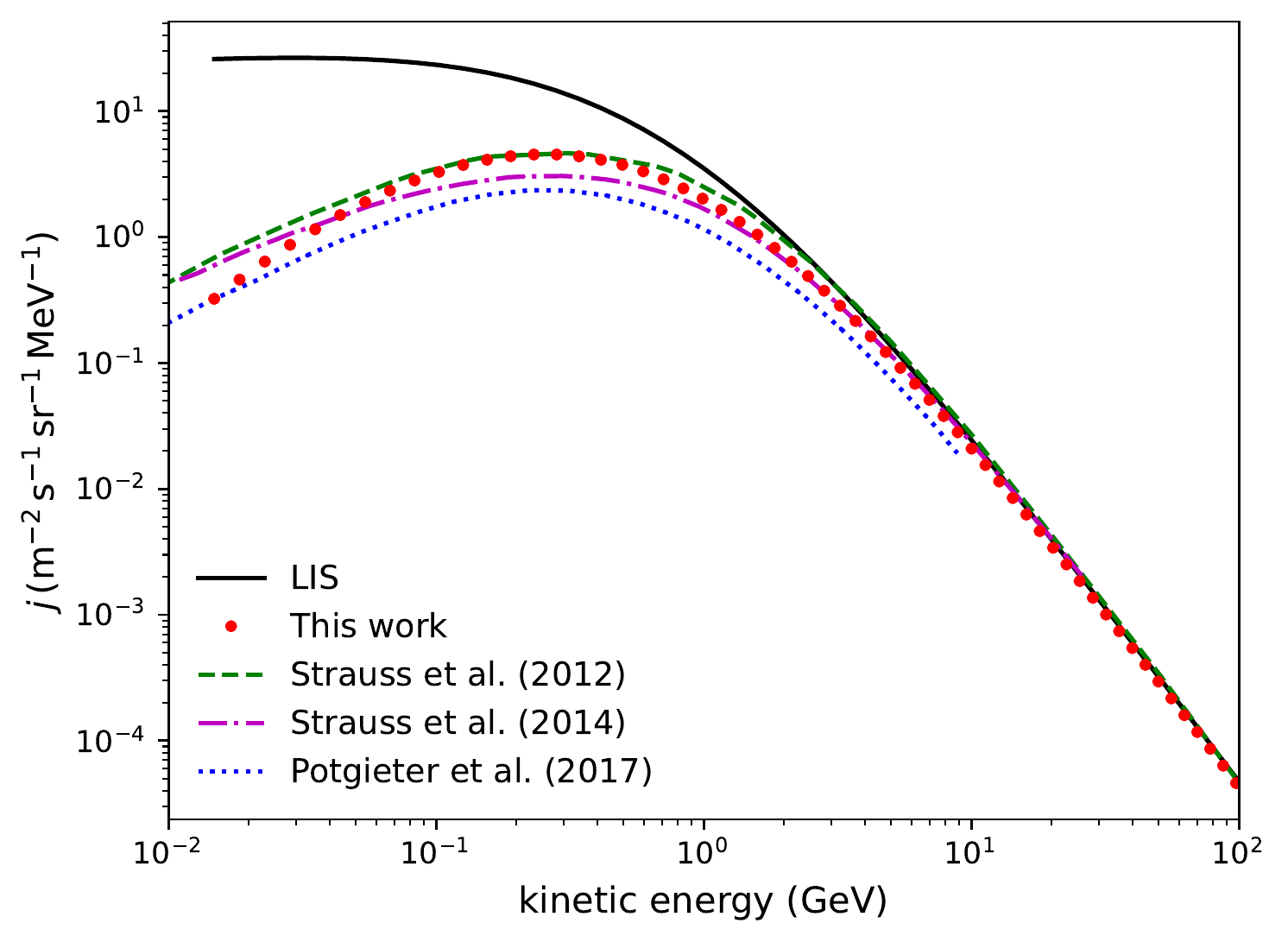}
    \caption{Differential intensity of Galactic cosmic rays as a function of kinetic energy at Earth representative of solar minimum conditions. The different linestyles represent different models for comparison, namely: red dots for this study, green dashed line for \citet{Strauss2014}, magenta dash-dotted line for \citet{Strauss2014} and blue dotted line for \citet{Potgieter2017}.}
    \label{fig:sun}
\end{figure}

Using this value, with the remaining Alfvén wave inputs being the same as those in \citet{Kavanagh2021}, we model the solar wind using a dipolar magnetic field with a polar strength of 3\,G, tilted 5$\degr$ with respect to the rotation axis. Our 3D spherical grid extends out to 20 solar radii, and contains around 4.5~million cells. The solar wind model used here is representative of solar minimum with the heliospheric magnetic field pointing outwards in the northern solar hemisphere (commonly referred to as A>0 cycles). The mass-loss rate we obtain from our model is $\sim3\times10^{-14}\,M_{\sun}~\textrm{yr}^{-1}$ and the solar wind terminal velocity is $540\,$km~s$^{-1}$ \citep[which is in the range of observed values, e.g.][]{McComas2000}. By extrapolating the magnetic field strength of the 3D model from its outer edge to 1\,au, we found a magnetic field strength at Earth's orbit of $\sim 4\,$nT \citep[which is in the range of observed values, e.g.][]{Potgieter2015}. We set the size of the heliosphere to 122\,au, as observed by {\it Voyager 1} \citep{Stone2013, Stone2019}. \Cref{fig:wind-sun} shows the solar wind velocity (left) and total magnetic field intensity (right) throughout the heliosphere. The total magnetic field 
resembles an $\sim$ aligned dipole, with a bimodal wind velocity distribution typical of solar wind at minimum \citep{McComas2000}. Using this model, we calculate the propagation of Galactic cosmic rays as discussed in \Cref{sec:crs}.The flux of cosmic rays found at Earth is shown as red dots in \Cref{fig:sun} as a function of the kinetic energy.

\Cref{fig:sun} shows the differential intensity of Galactic cosmic rays at Earth as a function of kinetic energy during the solar minimum activity and A>0 for different works, namely: this study (red dots), \citet{Strauss2014} (green dashed line), \citet{Strauss2014} (magenta dash-dotted line) and \citet{Potgieter2017} (blue dotted line). Overall, we get remarkably good agreement between our model and the 3D models to which we compare our results. The main discrepancies are most likely related to the different values of $B$, $u_\infty$ and heliospheric sizes used in each model and the fact that we use a 1D cosmic ray transport model while the other works use 3D models. 

Another point to note is that \citet{Potgieter2017} uses a `softening parameter' in the drift velocity equation so that the drift is reduced for cosmic rays with momentum below 0.55\,GeV/$c$. This reduction in the particle drift is necessary to explain the observations at Earth. In our models, we did not use a `softening parameter' to reduce the particle drift velocities to avoid including an extra free parameter in to our model.

\section{Extrapolation of stellar wind profiles to the edge of the astrosphere}
\label{sec:extrapolation}
The outer boundary of the stellar wind models for AU~Mic and Prox~Cen from \citet{Kavanagh2021} is at $100\,R_{\star}$, however the astrosphere extends much further out. Hence, we extrapolate the quantities $u$, $B_r$, $B_\theta$ and $B_\phi$ to account for the whole extent of the astrosphere. Beyond $100\,R_{\star}$, the radial velocity is extrapolated as a constant since the velocity has already reached its asymptotic value. To generate a Parker spiral \citep{Parker1958}, $B_r$ continues to fall with $r^{-2}$ and $B_\phi$ with $r^{-1}$. In the case of $B_\theta$, we use a power law fit for each system. For Prox~Cen, $B_\theta$ falls with $r^{-1.7}$, for `AU~Mic low' with $r^{-2.2}$ and for `AU~Mic high' with $r^{-4.7}$. For the Sun we extrapolate $B_\theta$ as $1/r^{4.6}$ beyond $r>0.04\,$au. \Cref{fig:wind,fig:wind-sun} show the extrapolated values of the wind velocity and total magnetic field for the stars studied here. Note, our stellar wind model does not take into account the termination shock (region where the stellar wind properties change due to the interactions with the local interstellar medium).

\bsp	
\label{lastpage}
\end{document}